\documentclass{article} 
\usepackage{mlspconf}

\copyrightnotice{978-1-7281-6662-9/20/\$31.00 {\copyright}2020 IEEE}

\usepackage[utf8]{inputenc}
\usepackage[T1]{fontenc}
\usepackage{subfig}
\usepackage[pdftex]{graphicx}
\usepackage{tikz}
\usetikzlibrary{spy}
\usepackage[colorlinks=true,linkcolor=blue,citecolor=blue]{hyperref}
\usepackage{url}
\usepackage{epstopdf}
\usepackage{amsmath,amssymb,amsfonts}
\usepackage{mathrsfs}
\usepackage{float}
\usepackage{cleveref}
\usepackage{enumitem}
\usepackage{multirow}
\usepackage{booktabs}
\usepackage{xcolor}
\usepackage[normalem]{ulem}

\DeclareMathOperator*{\argmax}{arg\,max}
\newcommand{\R}{\mathbb{R}}
\newcommand{\rev}[1]{{{#1}}}

\title{\LARGE \bf
Blind hierarchical deconvolution \\
}

\name{%
    A. Arjas$^{\star}$%
    \qquad L. Roininen$^{\dagger}$%
    \qquad M. J. Sillanp\"a\"a$^{\star}$
    \qquad A. Hauptmann$^{\star\diamond}$
    \thanks{This work was supported by the Academy of Finland (project no:s 312123, 326341, 334816, 334817) and by the Academy of Finland PROFI5 funding for mathematics 
    and AI: data insight for high-dimensional dynamics.
    }%
}
\address{%
    $^{\star}$ Research  Unit  of  Mathematical  Sciences,  University  of  Oulu,  Oulu  FI-90014 \\%
    $^{\dagger}$ School of Engineering Science, Lappeenranta-Lahti University of Technology, Lappeenranta FI-53851\\%
    $^{\diamond}$ Department  of  Computer  Science,  University College London, London WC1E 6BT, UK%
}

\begin{document}

\maketitle

\begin{abstract}

Deconvolution is a fundamental inverse problem in signal processing and the prototypical model for recovering a signal from its noisy measurement. Nevertheless, the majority of model-based inversion techniques require knowledge on the convolution kernel to recover an accurate reconstruction and additionally prior assumptions on the regularity of the signal are needed. To overcome these limitations, 
we parametrise the convolution kernel and prior length-scales, which are then jointly estimated in the inversion procedure. 
The proposed framework of \emph{blind hierarchical deconvolution} enables accurate reconstructions of functions with varying regularity and unknown kernel size and can be solved efficiently with an empirical Bayes two-step procedure, where hyperparameters are first estimated by optimisation and other unknowns then by an analytical formula.

\end{abstract}


\section{Introduction}

In machine learning, inverse problems, and signal processing, a typical problem is to make statistical estimation of an underlying signal given noisy convolved measurements.
Our objective is to deconvolve the signal given the noisy measurements, noise statistics and a convolution kernel function with an unknown kernel parameter.
This kind of a problem is called blind deconvolution \cite{haykin1994blind}.

Our proposed estimation algorithm is based on Bayesian statistical inference, where the solution is formed as a joint \emph{a posteriori} distribution for the unknown signal and the convolution kernel parameter. 
The posterior 
is proportional (up to a normalising constant) to the product of the likelihood and prior distributions, and in our case, we factorise our joint prior as a hierarchical model.
The choice of the denoising prior is based on various factors, such as the assumed smoothness, edges, or high-frequency components of the underlying signal. \rev{Consequently, it is inherently difficult to formulate a suitable prior for a large class of functions and is often approached by a combination of priors \cite{repetti2014euclid} to represent different features. In the following, we shall discuss the possibility to use a unifying prior by estimating the length-scale of the signal in a hierarchical model. } 

Gaussian processes are common choices for (denoising) priors, and they are covered in detail in the context of machine learning in Rasmussen and Williams 2006  \cite{Rasmussen2006} and in statistical inverse problems in Kaipio and Somersalo 2005 \cite{somersalokaipio}.
The recipe is rather simple, model your mean and covariance functions in the continuous time, and discretise them for practical computations.
However, the way you choose these functions is not trivial, and the choice dictates both inference accuracy, but also induces certain computational cost. 

For the sake of simplicity, let us choose a zero-mean Gaussian process (GP) prior with non-stationary covariance.
The construction of the non-stationarity covariance can be done in various ways -- some of the recent techniques include deep GPs in the sense of  Paciorek and Schervish 2006 \cite{Paciorek2006} and Damianou and Lawrence 2013 \cite{Damianou2013}. 
The analytical properties of deep GPs with applications to inverse problems was studied by Dunlop et al.\ 2018 \cite{Dunlop2018}, and a shallow two-layer alternative based on sparse presentations via stochastic partial differential equations (SPDE) was introduced in Roininen et al.\ 2019 \cite{Roininen2019}.
This was further developed by Monterrubio-G\'omez et al.\ 2020 \cite{monterrubio2020posterior}, where also MCMC techniques with elliptical slice sampling were developed for the extended model in \cite{Roininen2019}.

In this paper, we start by using the Paciorek and Schervish \cite{Paciorek2006} parameterisation of the covariance function, and model the length-scaling either as a Cauchy walk or a total variation (TV) prior, that is, the length-scaling is modelled as a non-Gaussian process.
TV priors correspond essentially to Laplace priors.
A similar construction was done for SPDE representation and Cauchy prior in \cite{Roininen2019}. 
We note that for full uncertainty quantification with respect to choice of the discretisation, Lassas and Siltanen 2004 \cite{Lassas2004} showed that under change of discretisation, the statistical estimators do not stay invariant. 
These can be alleviated by using Cauchy priors in the sense of Markkanen et al.\ 2019 \cite{Markkanen2019}. 
The reason TV priors are not discretisation-invariant is because they have finite moments, and every stochastic process with finite moments converges to a Gaussian process when the discretisation step goes to zero. 

We shall apply here the shallow 2-layer hierarchical model to blind deconvolution, where, on top of the unknown itself, we estimate the parameter of the convolution kernel. 
This leads to a severely ill-posed problem, and we show that with the proposed method of \emph{blind hierarchical deconvolution}, we can estimate the kernel parameter, and the unknown itself, which has smooth, linear and constant parts as well as sharp edges.
Typical prior choices limit often to one of these features, but our objective is to show that we can actually  construct models which are capable to recover all these features.

\section{A blind deconvolution model}
Deconvolution can be formulated as the basic linear inverse problem \cite{mueller2012linear} of recovering an unknown signal $\boldsymbol{f}\in \R^n$
from its noisy measurement $\boldsymbol{g} \in \R^n$, that is
\begin{equation}\label{eqn:linProb}
    \boldsymbol{g} = \boldsymbol{A}_\tau \boldsymbol{f} + \boldsymbol{e},
\end{equation}
where $\boldsymbol{e}\in \R^n$ denotes the noise component in the measurements and the forward mapping given as matrix $\boldsymbol{A}_\tau:\R^n \to \R^n$ models the relationship between signal and measurements. Typically, to obtain accurate reconstructions we assume to have full knowledge of the mapping $\boldsymbol{A}_\tau$. However, in the context of blind deconvolution, this is not the case and reconstruction quality is essentially limited by our ability to estimate the underlying convolution kernel.
In particular, we assume that the kernel depends on the parameter $\tau > 0$, which then needs to be jointly recovered with the actual signal for an accurate reconstructions. The full measurement model can be formulated as
\begin{equation}\label{eqn:convultionOp}
    (A_\tau f)(x) := (\phi(\cdot;\tau)\ast f)(x),
\end{equation}
with a Gaussian convolution kernel $\phi(x; \tau) = \frac{1}{\sqrt{2\pi\tau^2}}e^{-\frac{x^2}{2\tau^2}}$, where the parameter $\tau$ effectively controls the degree of blurring present in the measurements and is included as an unknown in the inversion process. In the following, $\boldsymbol{A}_\tau$ denotes the matrix representation of $A_\tau$ in \eqref{eqn:convultionOp} and  $\boldsymbol{f}$ is the discretisation of the continuous signal giving rise to the matrix-vector representation in \eqref{eqn:linProb}.

We emphasise that we consider here the prototypical problem of a Gaussian as convolution kernel, but other parameter dependent kernels\rev{, e.g. boxcar or triangular functions,}  can be considered as well in the following framework.

\subsection{Bayesian inversion}
Statistical or Bayesian inversion is a probabilistic framework for solving inverse problems. The solution to the problem is given as a probability distribution called the posterior distribution. 
The probability density function of the posterior distribution is obtained through the Bayes' formula
\begin{equation}
    p( \boldsymbol{f}, \boldsymbol{\theta}|\boldsymbol{g}) = \frac{p(\boldsymbol{g}| \boldsymbol{f}, \boldsymbol{\theta})p( \boldsymbol{f},\boldsymbol{\theta})
    }{p(\boldsymbol{g})},
\end{equation}
where $p(\boldsymbol{g}| \boldsymbol{f}, \boldsymbol{\theta})$ is the likelihood function that depends on the forward model and assumed distribution of the noise. Here a joint prior density $p( \boldsymbol{f},\boldsymbol{\theta})$ is hierarchically factored as $p( \boldsymbol{f},\boldsymbol{\theta})=p( \boldsymbol{f}|\boldsymbol{\theta})p(\boldsymbol{\theta})$, where $p( \boldsymbol{f}|\boldsymbol{\theta})$ is the prior density that encapsulates the prior assumptions concerning $\boldsymbol{f}$ given hyperparameters $\boldsymbol{\theta}$ and $p(\boldsymbol{\theta)}$ is the prior for hyperparameters; $p(\boldsymbol{g})$ is a normalising constant. The vector $\boldsymbol{\theta}$ consists of hyperparameters that are not interesting as such but control the properties of $\boldsymbol{f}$ and must be estimated with the help of the prior density $p(\boldsymbol{\theta})$.

Our empirical Bayes two-step approach for solving the inverse problem will be the following:
\begin{enumerate}[label=(\roman*)]
    \item Find the maximum a posteriori (MAP) estimate of the marginal posterior distribution of the hyperparameters, i.e. $\boldsymbol{\theta}_\text{MAP} = \argmax_{\boldsymbol{\theta}} p(\boldsymbol{\theta}|\boldsymbol{g})$. (Note that signal $\boldsymbol{f}$ is analytically integrated out from this expression).
    \item Find the closed form solution to the conditional posterior distribution of $\boldsymbol{f}$ given $\boldsymbol{\theta}_\text{MAP}$.
\end{enumerate}

\begin{figure*}[t!]
    \centering
    \includegraphics[width=.45\textwidth]{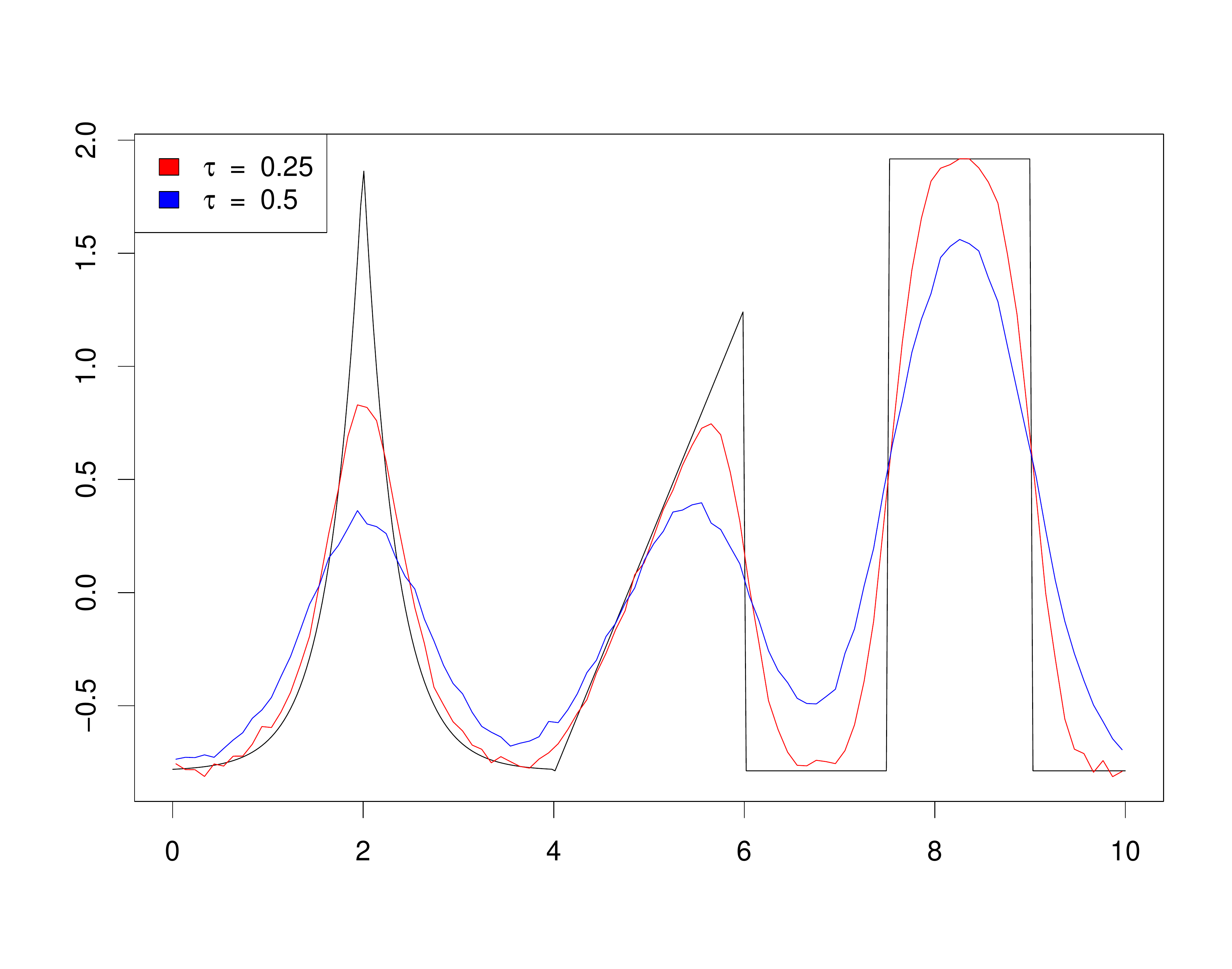}
    \includegraphics[width=.45\textwidth]{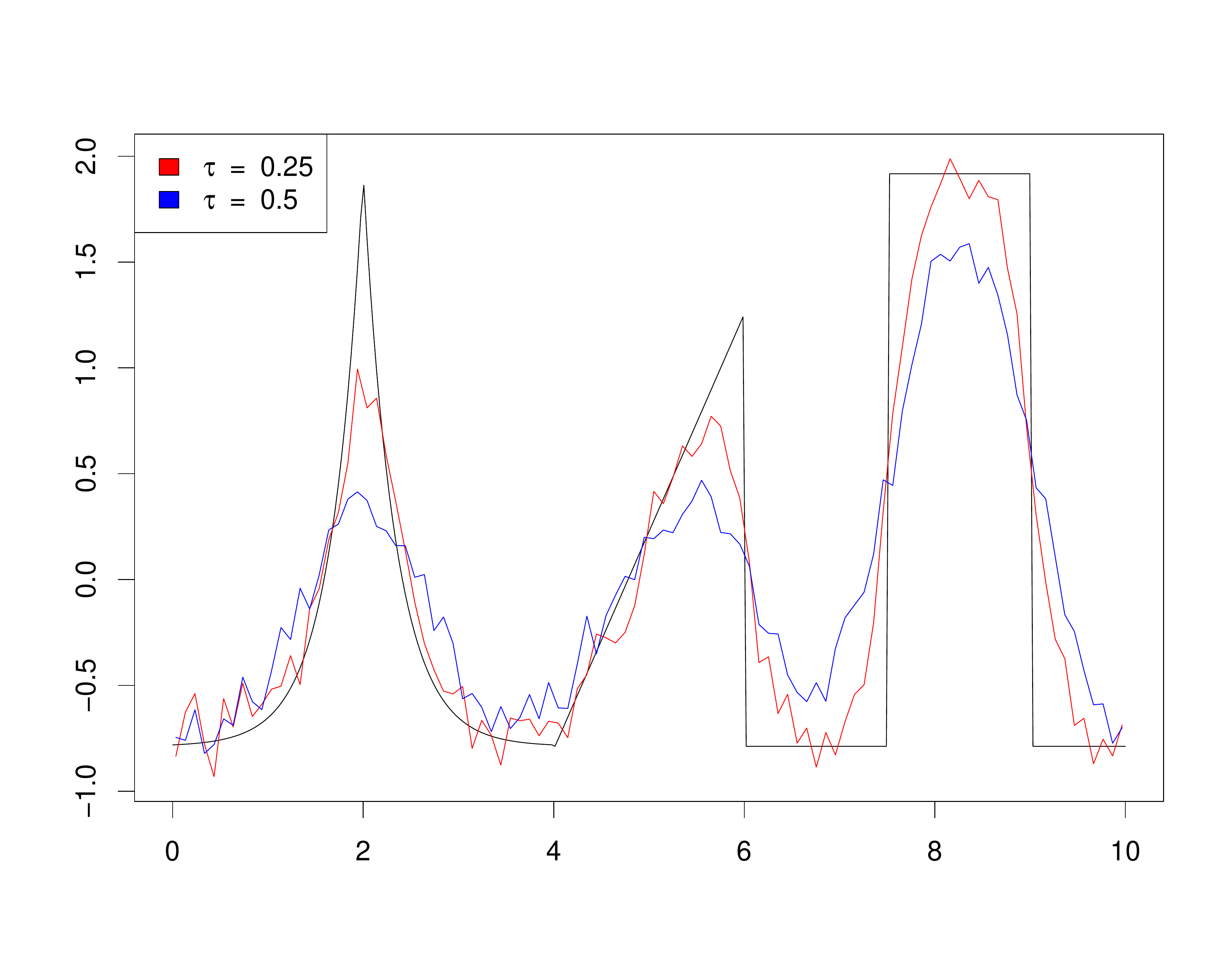}
    \caption{The unknown signal $\boldsymbol{f}$ drawn with black line and noisy measurements $\boldsymbol{g}$ \rev{(blue and red lines)} with 1\% noise (left) and with 5\% noise (right). The red lines indicate a convolution kernel with parameter $\tau=0.25$, the blue lines are for the wider convolution kernel with $\tau = 0.5$.    }
    \label{fig:meas}
\end{figure*}

\subsection{Prior models}
We construct different prior models for the unknown signal $\boldsymbol{f}$ and hyperparameters $\boldsymbol{\theta}$. Our goal is to be able to reconstruct both smooth and discontinuous/non-differentiable features of the signal. For this purpose, we introduce a length-scale function $\ell(\cdot)$. The idea is that when there are rapid changes or discontinuities in $\boldsymbol{f}$ at point $x$, the length-scale $\ell(x)$ is small, and consequently, when $\boldsymbol{f}$ is smooth or constant at $x$, $\ell(x)$ is large. We then incorporate this in the non-stationary Matérn covariance function \cite{monterrubio2020posterior}, defined as 
\begin{equation} \label{cns}
\begin{split}
    C^{\text{NS}}(x_i, x_j) = &\frac{\gamma^2(\ell(x_i)\ell(x_j))^{\frac{1}{4}}2^{1-\nu}}{\Gamma(\nu)L_{i,j}}\left(\frac{|x_i - x_j|}{L_{i,j}}\right)^\nu\\ &K_\nu\left(\frac{|x_i - x_j|}{L_{i,j}}\right), \\
    \text{with } L_{i,j}=&\sqrt{(\ell(x_i) + \ell(x_j))/2}
\end{split}
\end{equation}
where $x_i$ and $x_j$ are points in the domain of $\boldsymbol{f}$, $\gamma^2$ is a magnitude parameter, $\nu$ is a smoothness parameter and $K_\nu$ is the modified Bessel function of the second kind of order $\nu$. We fix all other parameters in the covariance function and estimate the length-scale function as a part of $\boldsymbol{\theta}$. We let $ \boldsymbol{f} \sim \mathcal{N}(\boldsymbol{0}, \boldsymbol{C}_{\ell}^\text{NS})$, where $\boldsymbol{C}_{\ell}^\text{NS}$ is a matrix whose entries consist of pairwise covariances between all measurement grid points calculated with Eq. \eqref{cns}.

We consider two different hyperprior models for the log length-scale function: Cauchy difference prior and total variation (TV) prior (essentially Laplace difference prior). These priors bring some stiffness to the length-scale function but the main idea is to improve its estimability by borrowing strength from adjacent covariate points with low computational cost. The priors can be formally expressed as
\begin{equation}\label{eqn:priors}
\begin{split}
    \log(\ell(x_{i})) - \log(\ell(x_{i - 1})) &\overset{\text{i.i.d.}}{\sim} \text{Cauchy}(0, \alpha), \text{ or}\\
    \log(\ell(x_{i})) - \log(\ell(x_{i - 1})) &\overset{\text{i.i.d.}}{\sim} \text{Laplace}(0, \alpha),
\end{split}
\end{equation}
where $\alpha$ acts as a regularisation parameter and must be chosen for each dataset individually and $i$ refers to the ith element of the vector. For the log convolution kernel width for the model \eqref{eqn:convultionOp}, we use a uniform prior: $\log(\tau) \sim U(-5, 0)$.

\subsection{Posterior distribution}
To help in calculation of the marginalised posterior distribution (i) and the conditional posterior distribution (ii) above, we will assume that $\boldsymbol{e} \sim \mathcal{N}(\boldsymbol{0}, \sigma^2\boldsymbol{I})$ with $ \boldsymbol{f} \perp \!\!\! \perp \boldsymbol{e}$. These assumptions allow for the analytic marginalisation of the posterior distribution over $\boldsymbol{f}$, yielding
\begin{equation}
    p(\boldsymbol{\theta}|\boldsymbol{g}) = \int_{\mathbb{R}^n}\frac{p(\boldsymbol{g}| \boldsymbol{f}, \boldsymbol{\theta})p( \boldsymbol{f}|\boldsymbol{\theta})p(\boldsymbol{\theta})}{p(\boldsymbol{g})}\mathrm{d}\boldsymbol{f} = \frac{p(\boldsymbol{g}|\boldsymbol{\theta})p(\boldsymbol{\theta})}{p(\boldsymbol{g})},
\end{equation}
where 
\begin{equation}
\begin{split}
    p(\boldsymbol{g}|\boldsymbol{\theta}) = &\frac{1}{\sqrt{(2\pi)^n\det(\boldsymbol{A}_\tau\boldsymbol{C}_{\ell}^{\text{NS}}\boldsymbol{A}_\tau^T + \sigma^2\boldsymbol{I})}}\\&\exp\left\{-\frac{1}{2}\boldsymbol{g}^T(\boldsymbol{A}_\tau\boldsymbol{C}_{\ell}^{\text{NS}}\boldsymbol{A}_\tau^T + \sigma^2\boldsymbol{I})^{-1}\boldsymbol{g}\right\}.
\end{split}
\end{equation}
The vector $\boldsymbol{\theta}$ consists now of the discretised length-scale function $\boldsymbol{\ell}$ and the convolution kernel width $\tau$. 

To obtain the final reconstruction of the signal, we note that the Gaussian assumptions also result in the Gaussianity of the conditional posterior of $\boldsymbol{f}$ given $\boldsymbol{\theta}_\text{MAP}$ with closed form solutions for the mean and covariance. That means, we have that $ \boldsymbol{f}|\boldsymbol{g}, \boldsymbol{\theta}_\text{MAP} \sim \mathcal{N}(\boldsymbol{\Bar{f}}_{\boldsymbol{\theta}_\text{MAP}}, \boldsymbol{\Bar{C}}_{\boldsymbol{\theta}_\text{MAP}})$ \cite{somersalokaipio}, where
\begin{equation}
\begin{split}
    \boldsymbol{\Bar{f}}_{\boldsymbol{\theta}_\text{MAP}} &= \boldsymbol{C}_{\widehat{\ell}}^{\text{NS}}\boldsymbol{A}_{\widehat{\tau}} ^T(\boldsymbol{A}_{\widehat{\tau}} \boldsymbol{C}_{\widehat{\ell}}^{\text{NS}}\boldsymbol{A}_{\widehat{\tau}}^T + \sigma^2\boldsymbol{I})^{-1}\boldsymbol{g},\\
    \boldsymbol{\Bar{C}}_{\boldsymbol{\theta}_\text{MAP}} &= \boldsymbol{C}_{\widehat{\ell}}^{\text{NS}} - \boldsymbol{C}_{\widehat{\ell}}^{\text{NS}}\boldsymbol{A}_{\widehat{\tau}}^T(\boldsymbol{A}_{\hat{\tau}} \boldsymbol{C}_{\widehat{\ell}}^{\text{NS}}\boldsymbol{A}_{\widehat{\tau}}^T + \sigma^2\boldsymbol{I})^{-1}\boldsymbol{A}_{\widehat{\tau}} \boldsymbol{C}_{\widehat{\ell}}^{\text{NS}}.
\end{split} 
\label{eqn:Postform}
\end{equation}
The hat-notation indicates that the matrices are constructed with the MAP estimates of the parameters.

\section{Computational experiments}

We will test the performance of the proposed framework of \emph{blind hierarchical deconvolution} with a set of computational experiments. In order to assess the capability to reconstruct functions of varying regularity we chose the ground-truth signal (Fig. \ref{fig:meas}) to consist of a smooth exponential spike, a linear ramp as well as a piece-wise constant. 
To avoid inverse crime\footnote{\rev{Inverse crime is understood as making the inversion too easy by using the same discretisation for generating the data and the inversion process.}}, we simulated the data on a grid of 300 equidistant points and downsampled the measurements to the final grid size of 100 equidistant points. This way, we make sure that the inversion process is more challenging and represents a realistic scenario, where discretisation errors are inherently present. 

The discrete signal is convolved with a Gaussian convolution kernel following \eqref{eqn:convultionOp} with two different kernel widths by choosing $\tau = 0.25$ and $\tau = 0.5$. We have then added Gaussian noise to both convolved signals with relative noise level of 1\% and 5\%, yielding in total four datasets of varying difficulty.

To compute the MAP estimate of the hyperparameters, we used a limited memory BFGS algorithm \cite{bfgs}, implemented in the R function \texttt{optim()}. To avoid errors caused by numerical inaccuracies, we constrained the parameters to a maximum of $\text{log}(1000)$. For the parameters of the Matérn covariance function \eqref{cns}, we set $\gamma^2 = 1$ and $\nu = 1.5$. The final reconstruction $\boldsymbol{\Bar{f}}_{\boldsymbol{\theta}_\text{MAP}}$ is then computed with equation \eqref{eqn:Postform}.

\begin{figure}[t!]
    \centering
    \includegraphics[width=.5\textwidth]{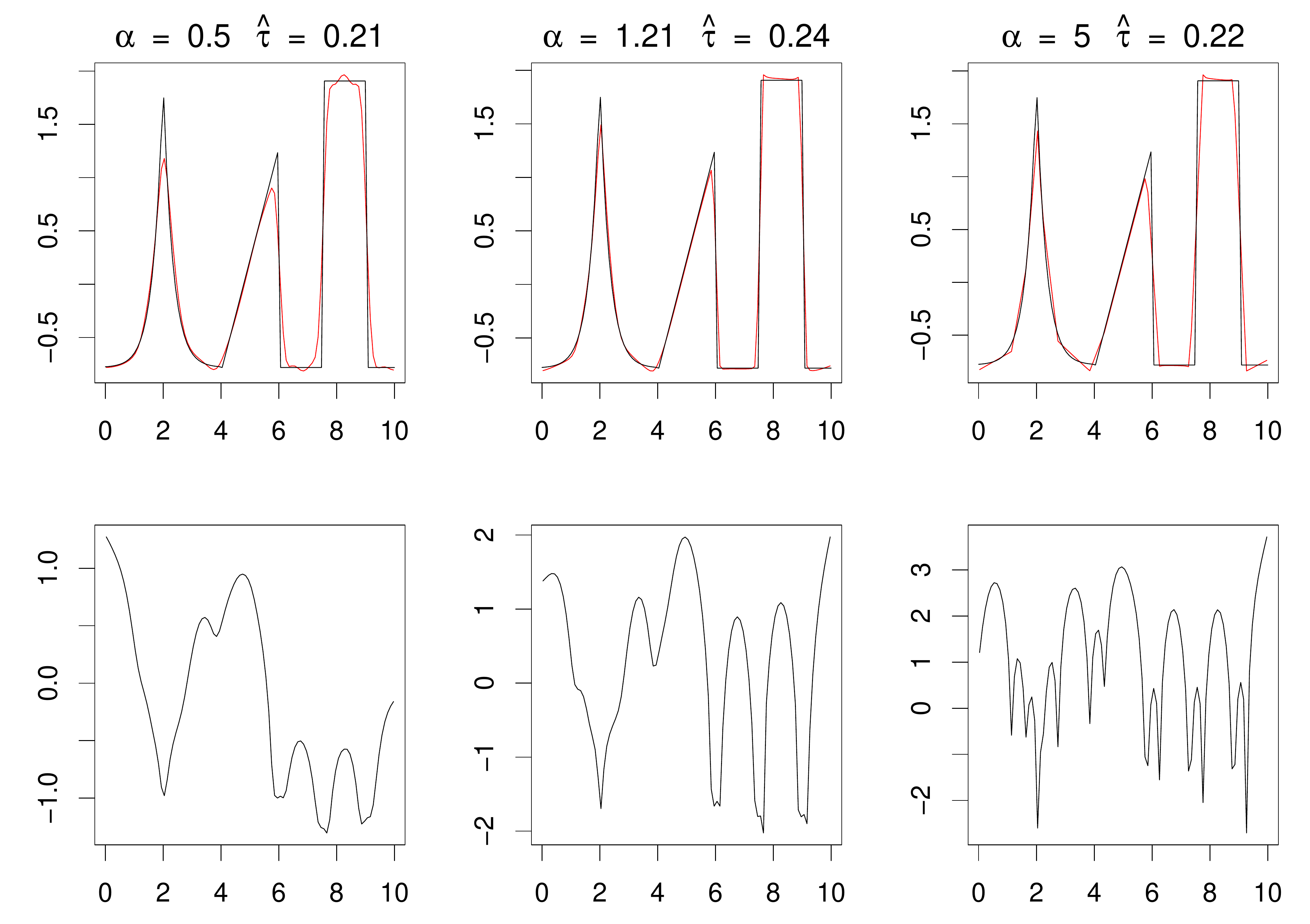}
    \caption{Results with the Cauchy difference prior for 1\% noise and $\tau = 0.25$ with different values of $\alpha$ indicated on top of the figures. Additionally, we present the estimated parameter $\hat{\tau}$ for each case. Top row: reconstruction of the unknown signal $\boldsymbol{f}$. Bottom row: estimates of the logarithmic length-scale function.}
    \label{fig:c1}
\end{figure}

\begin{figure}[t!]
    \centering
    \includegraphics[width=.5\textwidth]{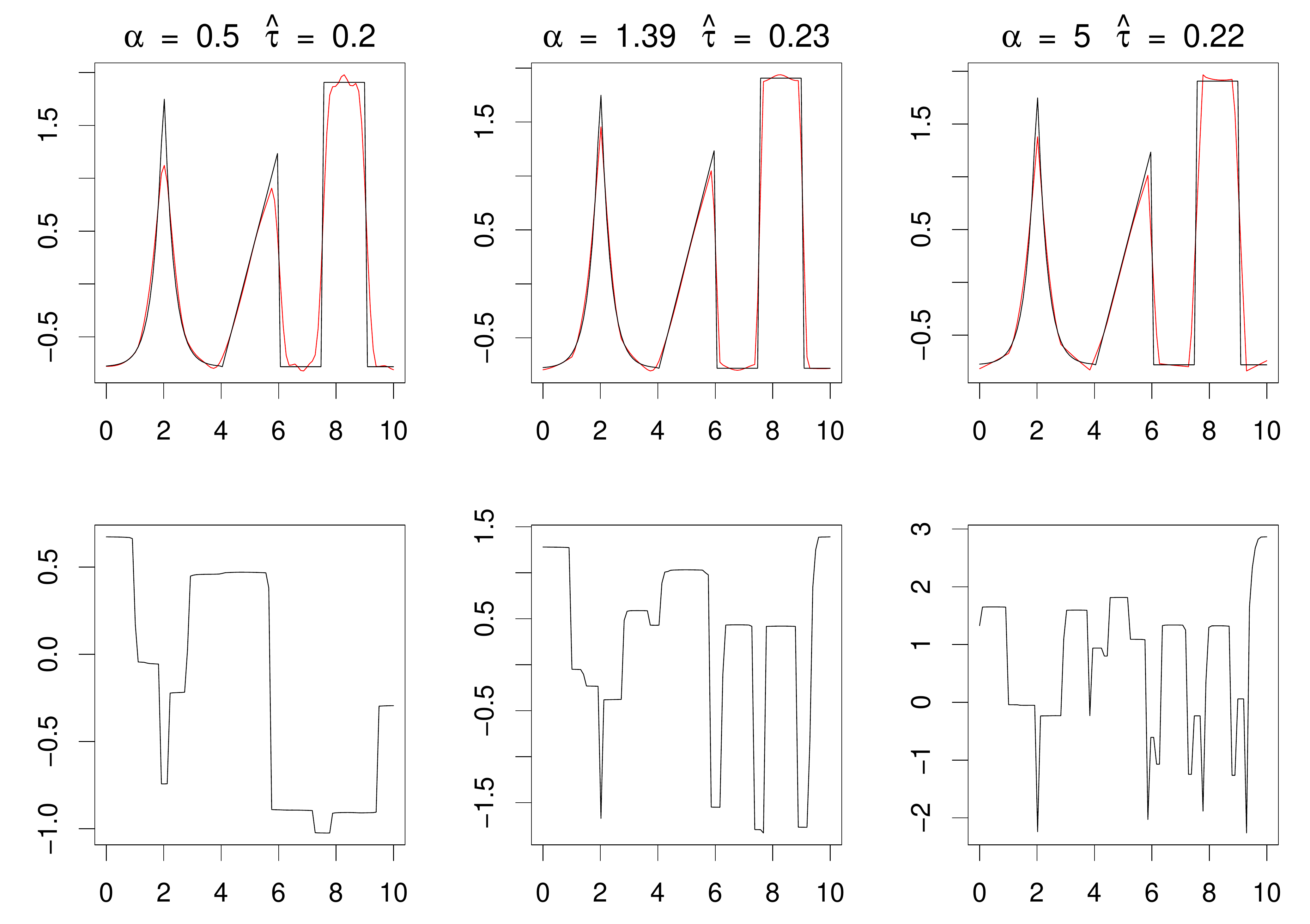}
    \caption{Results with the TV prior for 1\% noise and $\tau = 0.25$ with different values of $\alpha$ indicated on top of the figures. Additionally, we present the estimated parameter $\hat{\tau}$ for each case. Top row: reconstruction of the unknown signal $\boldsymbol{f}$. Bottom row: estimates of the logarithmic length-scale function.}
    
    \label{fig:TV1}
\end{figure}

An important aspect in the reconstruction is the choice of regularisation parameter $\alpha$ for the priors in \eqref{eqn:priors}, which will influence the reconstructed characteristics. To find such an optimal regularisation parameter $\alpha$, we performed a grid search for each dataset and prior, to find the parameter such that the mean squared error between the estimate of $\boldsymbol{f}$ and ground truth was minimised. \rev{This method of choosing $\alpha$ is clearly unsuitable for measured signals, where the ground truth is not available and hence secondary indicators would be needed to choose a suitable regularisation parameter. Nevertheless, we concentrate here on demonstrating the potential of the proposed framework and in order to illustrate the} importance of this choice we present results with too small and too large value of the regularisation parameter $\alpha$ in Figure \ref{fig:c1} and \ref{fig:TV1}.

 The total reconstruction times for the Cauchy prior range between 1 and 8 minutes. For the TV prior this increases to 5 to 16 minutes. In our observations, this indicates that convergence to the MAP estimate of the hyperparameters is harder to achieve with the TV prior which is likely due to the non-differentiability of the Laplace probability density function.
In the following we will qualitatively and quantitatively evaluate the performance of the proposed model.

\subsection{Comparison to stationary Gaussian Process model}
To demonstrate the effectiveness of the non-stationary covariance function, we also present results using a stationary covariance function. This means that the length scaling $\ell(\cdot)$ in expression \eqref{cns} is reduced to a scalar. This kind of a model is easier and faster to use but overfits if the length-scale is too low and oversmooths the edges when it is too high. We fit the model similarly to the non-stationary model, optimising the hyperparameters and using the expression \eqref{eqn:Postform} to reconstruct the signal.

\section{Discussion}
\begin{figure}[t!]
    \centering
    \includegraphics[width=.5\textwidth]{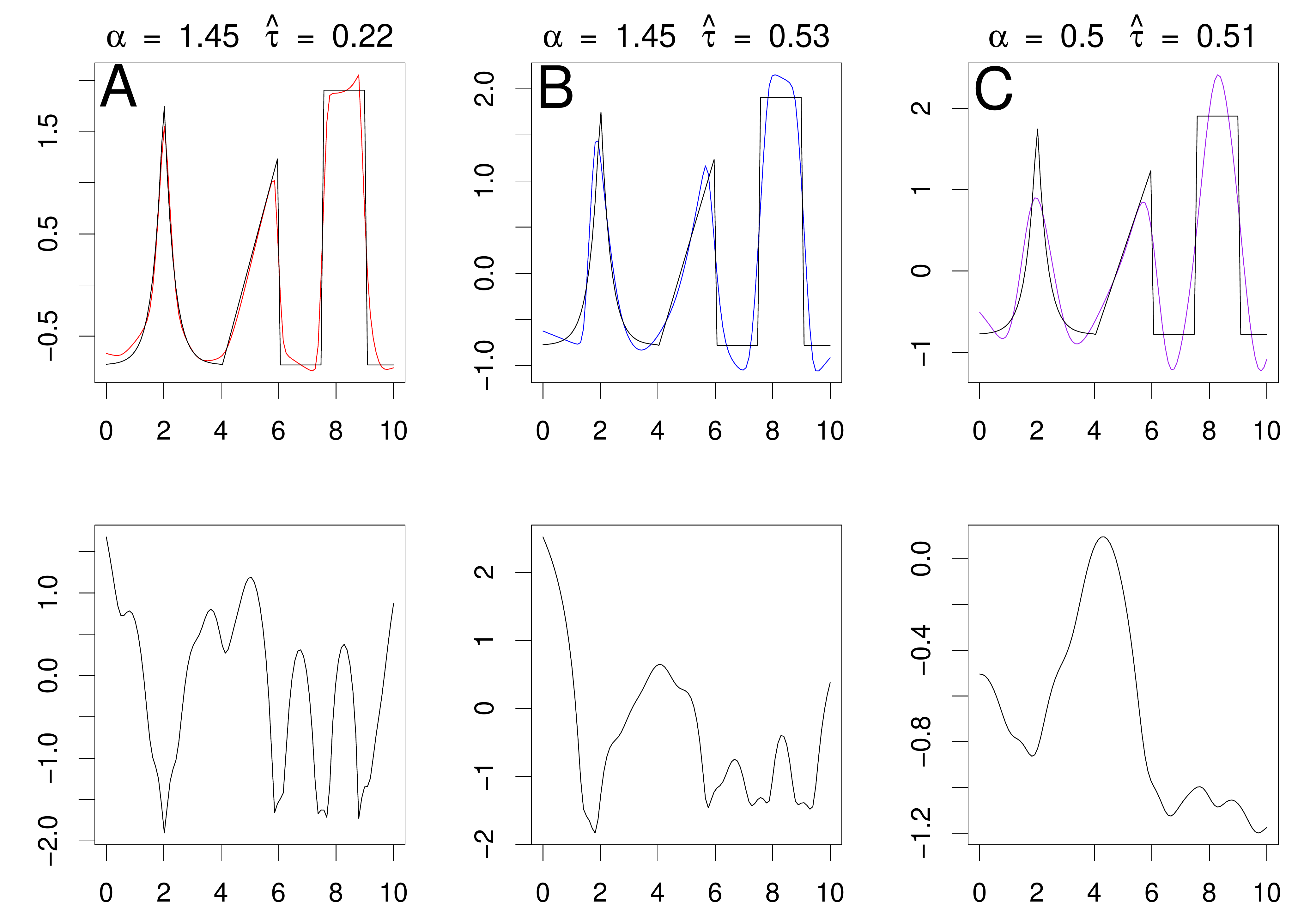}
    \caption{Estimates of $\boldsymbol{f}$ with the with Cauchy difference prior for the simulations with A: 5\% noise and $\tau = 0.25$, B: 1\% noise and $\tau = 0.5$ and C: 5\% noise and $\tau = 0.5$ . Bottom: estimates of the corresponding logarithmic length-scale function.}
    \label{fig:c2}
\end{figure}

\begin{figure}[t]
    \centering
    \includegraphics[width=.5\textwidth]{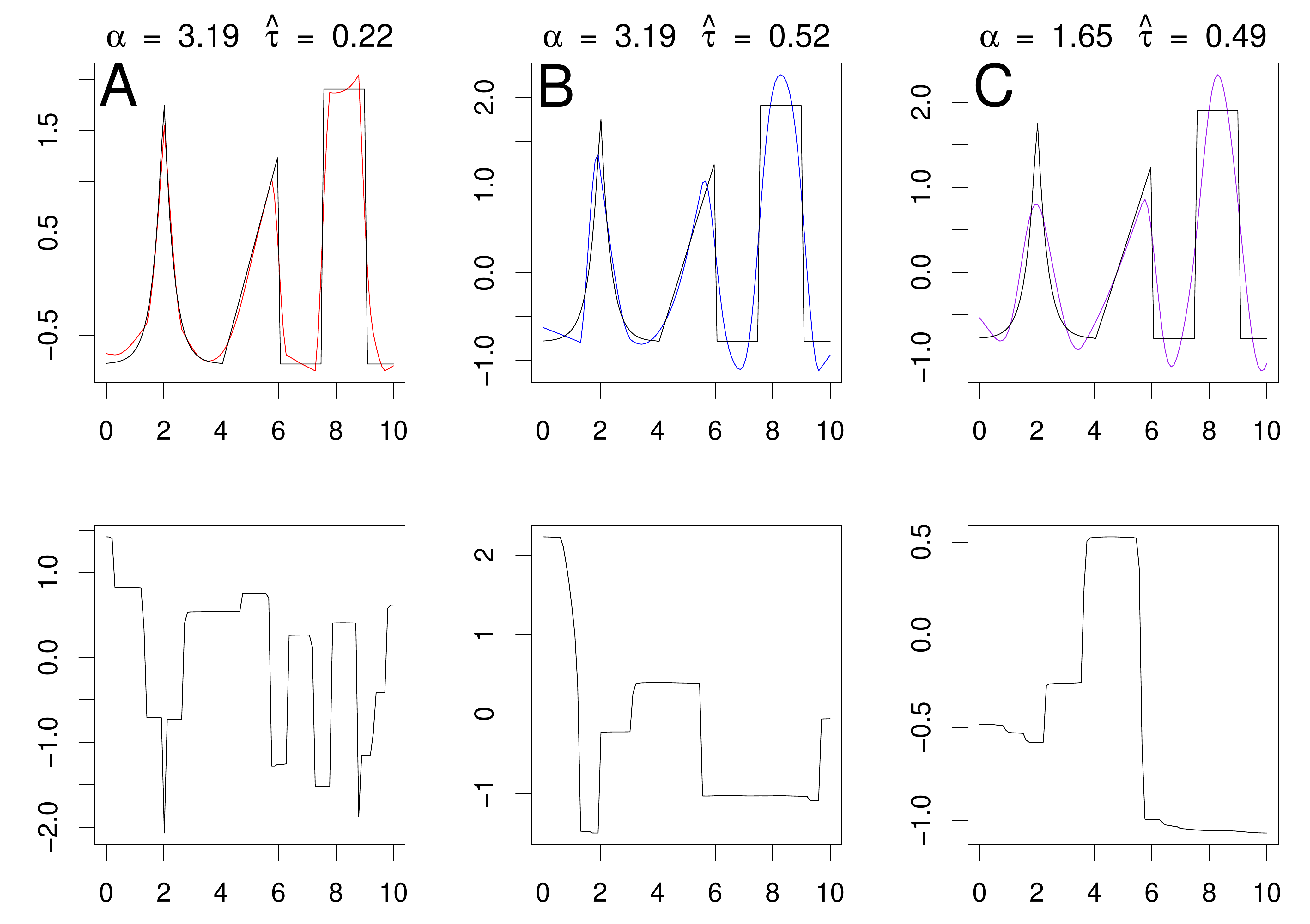}
    \caption{
    Estimates of $\boldsymbol{f}$ with the with TV prior for the simulations with A: 5\% noise and $\tau = 0.25$, B: 1\% noise and $\tau = 0.5$ and C: 5\% noise and $\tau = 0.5$ . Bottom: estimates of the corresponding logarithmic length-scale function.}
    \label{fig:TV2}
\end{figure}

Let us first discuss the visual performance of our proposed method. The reconstructions for the datasets with the Cauchy difference prior are presented in Figures \ref{fig:c1} and \ref{fig:c2}, for the TV prior in Figures \ref{fig:TV1} and \ref{fig:TV2}. 
In general we can say, that the model performs excellent in the case with low noise and a narrow convolution kernel, but it deteriorates clearly for more difficult cases, as is expected.

With both priors, in the case of the convolution kernel with $\tau=0.25$, 1\% noise, and an optimal choice of regularisation parameter, we are able to estimate the convolution kernel with a slight offset as \rev{$\hat{\tau} = 0.24$ for the Cauchy prior and $\hat{\tau} = 0.23$ for the TV prior}. The reconstructed signals, shown in Figure \ref{fig:c1} for the Cauchy prior and Figure \ref{fig:TV1} for the TV prior, successfully recover the varying characteristics of the unknown. The smooth parts are nicely recovered, as well as the linear ramp and the spike. 
The behaviour of the length-scale estimates is as desired, since they descend rapidly to recover the sharp edges and ascends for the smoother parts. We can see that in the case of the TV prior, we get more piece-wise constant estimates than in the Cauchy case, which is typical for TV priors. 
This indicates, that the Cauchy prior is better suitable to recover the smoother parts of the signal as it is more flexible with respect to varying length-scales, whereas for the TV prior the changes are more sudden and hence does not favour smooth changes in the length-scale. Overall, we can say that the performance is excellent for this case.


For the data with higher noise and wider convolution kernel results deteriorate clearly, as shown in Figure \ref{fig:c2} and \ref{fig:TV2}. If we only increase the noise to 5\%, reconstruction results are still comparably good and in particular, our model is able to estimate the convolution parameter \rev{with $\hat{\tau}=0.22$} for both prior models. Consequently, features of the ground-truth signal are still rather well preserved in the reconstructions. For the wider convolution kernel with $\tau=0.5$, our model has difficulties to overcome the loss of information with any choice of the regularisation parameter and the estimates of the convolution kernel become less accurate. For lower noise, the Cauchy prior performs slightly better, as it can still capture the changing regularity in the signal. For 5\% noise, both priors tend to provide too smooth reconstructions and are not able to estimate the length-scales properly anymore.

 

\begin{figure}[t!]
    \centering
    \includegraphics[width=.5\textwidth]{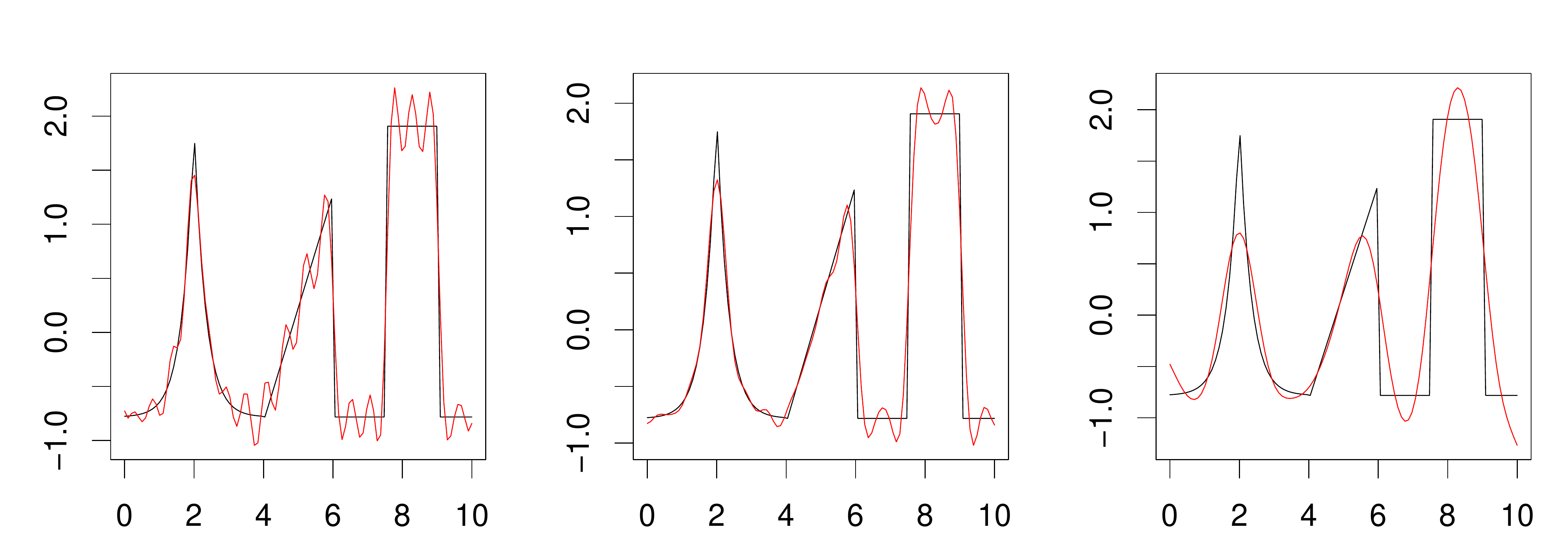}
    \caption{Estimates of $\boldsymbol{f}$ with a stationary Gaussian process for the data with 1\% noise and $\tau = 0.25$ with too short (left), optimal (middle) and too long (right) length-scaling.}
    \label{fig:statGP}
\end{figure}

For comparison, results for the stationary Gaussian process are presented in Figure \ref{fig:statGP} for the data with 1\% noise and $\tau = 0.25$. Too short length-scale overfits the noise and too long length-scale oversmooths the edges. The optimal length-scale is a compromise between edgy and smooth features. The edges are somewhat smoothed out and the constant and linear parts are reconstructed as wavy. Hence, adaptive length-scaling is clearly beneficial for this kind of signal. The convolution kernel width is estimated correctly as $\hat{\tau} = 0.25$.

Finally, we present quantitative values in Table \ref{table:quantResults} measured as relative MSE, which confirm the visual results that the Cauchy prior does perform generally better to recover the true signal. For $\tau=0.25$ both methods do perform well in recovering the unknown signal with a relative reconstruction error of less than $5\%$ for the low noise case and slightly higher values for the high noise case. These values considerably decrease for the wider convolution kernel, confirming our visual evaluation. Here, the influence of noise amplitude has a much larger impact on the reconstruction quality. Finally, we note that the relative MSE for the stationary Gaussian process reconstruction, with only 1\% noise, is 6.1\% which is clearly larger than for the non-stationary Gaussian process.

\begin{table}[t!] 
\small
  \caption{Quantitative measures for the obtained reconstructions in relative MSE (in \%) with respect to the ground-truth $\boldsymbol{f}$ for all cases under considerations and an optimal choice of regularisation parameter.}
\begin{tabular}{lcc}
 & & \\ 
\multicolumn{3}{c}{Cauchy prior}\\
\hline
 & \multicolumn{2}{c@{}}{$\tau$}\\
\cmidrule(l){2-3}
Noise $\sigma^2$ & 0.25 &  0.5\\ 
\midrule
  1\%  & 3.8\% &  10.9\% \\
  5\%  & 6.1\% &  16.9\%\\
  \label{error}
\end{tabular}
\qquad
\begin{tabular}{lcc}
\multicolumn{3}{c}{TV prior}\\
\hline
 & \multicolumn{2}{c@{}}{$\tau$}\\
\cmidrule(l){2-3}
Noise $\sigma^2$ & 0.25 &  0.5\\ 
\midrule
  1\%  & 4.0\% &  12.9\% \\
  5\%  & 7.0\% &  16.7\%\\
\end{tabular}

  \label{table:quantResults}%
\end{table}%

\section{Conclusion}

In this work we have discussed the possibility to perform blind deconvolution of a measured signal with varying regularity. In the presented model, we assume a parameter dependency of the convolution kernel, which then can be estimated jointly with a non-stationary length-scale function to enable reconstructions of varying regularity.

We evaluated the performance of the proposed framework on a series of experiments with increasing ill-posedness and  showed that we are able to successfully estimate the convolution kernel from the measured data only. The estimated length-scale function is capable to adjust to the different features in the ground-truth signal, from smooth to linear and constant parts. Thus, we can recover signals of varying regularity with one prior only. We observed, that the Cauchy difference prior performs better in our experiments, than the TV prior. Nevertheless, both priors were not able to recover a meaningful signal anymore from strongly convolved and high noise signals. More work needs to be done in this case to compensate for such extreme scenarios.

As the computation of solutions is optimisation based, the proposed methodology should be directly applicable to high-dimensional data-intensive machine learning problems, but also to spatial statistics and inverse problems with large parameter space, such as in imaging applications.
In future studies, the applicability to uncertainty quantification, in the sense of \cite{monterrubio2020posterior}, should also be carried out while maintaining computational efficiency.

\bibliographystyle{IEEEbib}
\bibliography{refs}

\end{document}